\documentclass[a4paper]{mem}
\usepackage{natbib}
\usepackage{graphicx}
\usepackage[a4paper]{hyperref}
\newcommand{\rxa}{\hbox{RX\,J0420.0$-$5022}}
\newcommand{\rxb}{\hbox{RX\,J0720.4$-$3125}}
\newcommand{\rxc}{\hbox{RX\,J0806.4$-$4123}}
\newcommand{\rxd}{\hbox{1RXS\,J130848.6+212708}}
\newcommand{\rxe}{\hbox{RX\,J1605.3+3249}}
\newcommand{\rxf}{\hbox{RX\,J1856.4$-$3754}}
\newcommand{\rbs}{\hbox{RBS1223}}
\newcommand{\pdot}{\dot P}
\begin{document}
   \title{The XMM-Newton view of radio-quiet and X-ray dim isolated neutron stars}

   \author{F. Haberl}

   \offprints{F. Haberl, \email{fwh@mpe.mpg.de}}

   \institute{Max-Planck-Institut f\"ur extraterrestrische Physik,\\
           Giessenbachstra{\ss}e, 85748 Garching, Germany}

\abstract{
As part of the guaranteed time, guest observer and calibration programs, XMM-Newton 
extensively observed a group of six thermally emitting, isolated neutron stars 
which are neither connected with a supernova remnant nor show 
pulsed radio emission. The high statistical quality of the EPIC data allows 
a detailed and homogeneous analysis of their temporal and spectral properties.

Four of the six sources are now well established as X-ray pulsars:
The 11.37 s period discovered in EPIC-pn data of \rxc\ was confirmed 
in a second XMM-Newton observation. In the case of the X-ray 
faintest of the six stars, \rxa\, the period marginally indicated in
ROSAT data was not seen in the EPIC data, instead
a 3.45 s pulse period was clearly detected.
Spectral variations with pulse phase were discovered for the known 8.39 s 
pulsar \rxb\ and also \rbs. For the latter EPIC data revealed a 
double-peaked pulse profile for a neutron star spin period of 10.31 s. 

The X-ray continuum spectra of all six objects are consistent with
a Planckian energy distribution with black-body temperatures kT between 40 eV and 
100 eV. EPIC data of the pulsars \rbs\ and \rxb\ revealed 
a broad absorption feature in their spectra at energies of 100-300 eV 
and $\sim$260 eV, respectively. The depth of this feature varies with pulse phase
and may be caused by cyclotron resonance scattering of protons or heavy 
ions in a strong magnetic field. In such a picture the inferred field strength 
exceeds 10$^{13}$ Gauss, in the case of \rxb\ consistent with the 
value estimated from its pulse period derivative. A similar absorption 
feature in the RGS and EPIC spectra of \rxe\ was reported recently.

\keywords{X-rays: stars -- stars: neutron -- stars: magnetic fields}}
\authorrunning{F. Haberl}
\titlerunning{Radio-quiet and X-ray dim isolated neutron stars}
\maketitle

\section{Introduction}

   \begin{figure*}
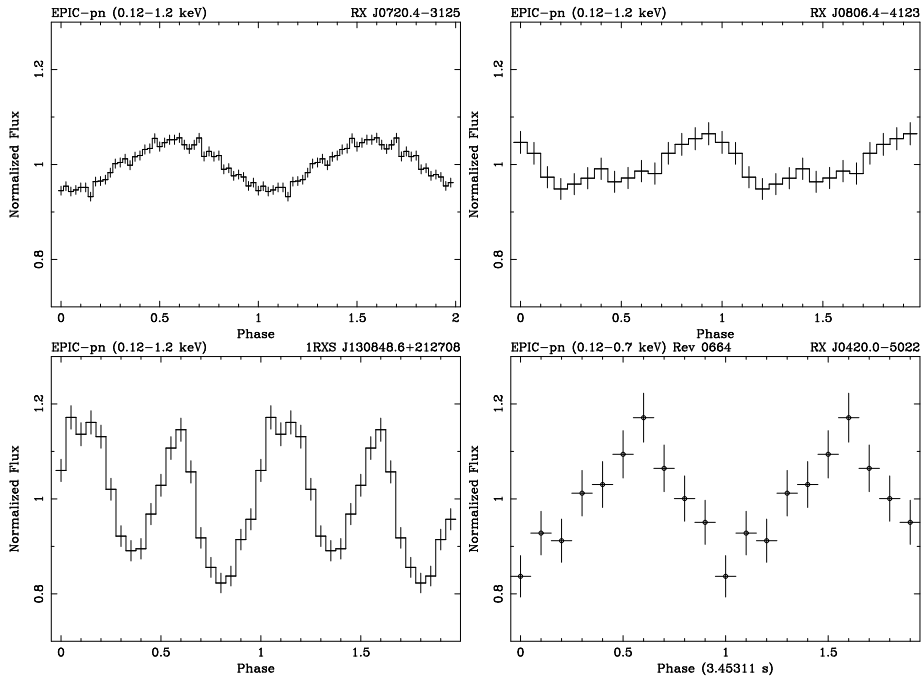

   \centering
   \includegraphics[angle=-90,width=6cm]{fhaberl1a.ps}
   \includegraphics[angle=-90,width=6cm]{fhaberl1b.ps}
   \vspace{3mm}
   \includegraphics[angle=-90,width=6cm]{fhaberl1c.ps}
   \includegraphics[angle=-90,width=6cm]{fhaberl1d.ps}
      \caption{EPIC-pn light curves folded on the pulse period for the
               four pulsars among the XDINs. 
              }
         \label{fig-puls}
   \end{figure*}

Presently more than half a dozen ROSAT-discovered X-ray dim isolated neutron stars 
\citep[XDINs, for recent reviews see][]{2000PASP..112..297T,2001xase.conf..244M,haberl2002COSPAR}
are known which share very similar properties. Their X-ray spectra are characterized 
by soft blackbody-like emission without indication for harder, non-thermal 
components. These stars apparently show no pulsed radio emission and no 
association with supernova remnants. Some of them exhibit pulsations 
in their X-ray flux indicating the neutron star rotation period.
Although there is little doubt that the soft X-rays are of thermal origin  
from the surface of an isolated neutron star, the details for the formation 
of the spectrum are not clear.
XMM-Newton observed six XDINs as part of the guaranteed 
time, guest observer and calibration programs. Here I summarize some of the 
first results of a homogeneous analysis of their temporal and spectral properties. 

\section {X-ray pulsations}

The second brightest of the known XDINs is \rxb\ and was the first discovered as
X-ray pulsar \citep{1997A&A...326..662H} with a period of 8.39 s. XMM-Newton 
observed \rxb\ six times. As example the folded EPIC-pn light curve from
satellite revolution 534 is drawn in Fig.~\ref{fig-puls} which shows a pulsed fraction 
of about 11\%. In Chandra data of \rbs\ = \rxd\ \citet{2002A&A...381...98H} discovered pulsations
with a period of 5.16 s and XMM-Newton observations revealed that the true neutron 
star spin period is more likely 10.31 s. This is supported by pulse phase dependent 
hardness ratio variations which are different for the two intensity maxima 
\citep{2003A&A...403L..19H}. \rbs\ exhibits the deepest modulation of the known XDINs 
in its double peaked pulse profile of 18\% (Fig.~\ref{fig-puls}). The first XMM-Newton 
observation of \rxc\ revealed a 6\% modulation (Fig.~\ref{fig-puls}) with a period of 11.37 s 
\citep{2002A&A...391..571H} which was confirmed in a second observation \citep{2004A&A.2Haberl}.
Four XMM-Newton observations of \rxa\ did not confirm the 22.7 s pulsations originally indicated 
in ROSAT data, but clearly reveal a 3.45 s period. The pulsed fraction is about 12\% 
(Fig.~\ref{fig-puls}). In Table~\ref{tab-sum} the derived pulse periods 
for the four pulsars are summarized together with pulse period derivatives assuming linear
period changes between multiple XMM-Newton observations. In most cases the time base
line is too short to determine precise $\pdot$ values, only for \rxb\ observations by different
satellites cover already more than 10 years \citep{2002MNRAS.334..345Z,2002ApJ...570L..79K}.

\section {X-ray spectra}

The X-ray spectra of XDINs obtained by the ROSAT PSPC were all consistent with black-body emission
little attenuated by interstellar absorption suggesting that the objects are close-by. To look 
for absorption features which may be created 
by heavy chemical elements in the stellar atmosphere high resolution spectra of \rxf\ were
obtained by the LETGS on Chandra \citep{2001A&A...379L..35B,2003A&A...399.1109B}.
Surprisingly, no significant narrow features were detected.

A spectral analysis (in a homogenous way using the latest calibration data) of the EPIC-pn spectra,
which are of unprecedented statistical quality, shows that in several cases a black-body model 
yields unsatisfactory fits. The strongest deviations are seen from \rbs\ and \citet{2003A&A...403L..19H}
demonstrate that a non-magnetic atmosphere models can neither explain the observed spectrum.
However, it was found that adding an absorption feature modeled by a broad (100 eV) Gaussian line at
an energy between 100 and 300 eV to the Planckian continuum yields acceptable fits. 
Similarly, but at a higher energy of 450 eV and therefore inside the sensitive band of the RGS instruments
\citet{2004vanKerkwijk} presented the detection of a broad absorption feature in the spectra of \rxe.
\citet{2004A&A.Haberl} report a broad absorption feature in the EPIC-pn spectra of the pulsar \rxb\ 
at an energy of 270 eV and find that the depth of the feature varies with pulse phase by a factor of $\sim$2.
Finally, also spectra of \rxa\ indicate a possible absorption line at 330 eV \citep{2004A&A.2Haberl}.

In Table~\ref{tab-sum} the spectral parameters inferred from the fits using the two models (absorbed black-body and 
absorbed black-body with Gaussian-shaped absorption line) are listed. The first value given for column density N$_{\rm H}$
and black-body temperature kT refers to the model without absorption line and the second to the model with line added.

\begin{table*}
\caption{X-ray-dim isolated neutron stars observed by XMM-Newton}
\begin{tabular}{lcccccc}
%\noalign{\smallskip}
\hline\noalign{\smallskip}
  Source               & $P$     & $\pdot$                & N$_{\rm H}$         & kT        & E$_{\rm line}$ & B \\
                       & (s)     & (s~s$^{-1}$)           & $10^{20}$ cm$^{-2}$ & (eV)      & (eV)           & 10$^{13}$ G  \\
\noalign{\smallskip}\hline\noalign{\smallskip}
RX\,J0420.0$-$5022     &  3.453  & $<9\times10^{-12}$     & 1.0/2.0             & 45/45     & 330?           & 6?  \\
RX\,J0720.4$-$3125     &  8.391  & $(3-6)\times10^{-14}$  & 1.4/0.8             & 85/84     & 270            & 5   \\
RX\,J0806.4$-$4123     & 11.371  & $<2\times10^{-12}$     & 0.4                 & 96        & $-$            &     \\
RBS1223                & 10.313  & $<6\times10^{-12}$     & 7.1/4.1             & 95/86     & 100-300        & 2-6 \\
RX\,J1605.3+3249       &   $-$   & $-$                    & 0.3/0.9             & 96/93     & 450            & 9   \\
RX\,J1856.5$-$3754     &   $-$   & $-$                    & 0.9                 & 60        & $-$            & 1?  \\
\noalign{\smallskip}\hline\noalign{\smallskip}
\end{tabular}
\label{tab-sum}
\end{table*}

\section{Discussion}

Four of the six X-ray-dim isolated neutron stars are now known as X-ray pulsars with spin periods between
3.45 s and 11.37 s. The fraction of pulsed flux in their folded X-ray light curves ranges between 6\% and 
18\%. \rbs, \rxb\ and \rxa\ show hardness ratio variations with pulse phase 
\citep{2003A&A...403L..19H,2004A&A.Haberl,2004A&A.2Haberl} while for \rxc\ the shallow modulation makes the 
significant detection of such an effect more difficult \citep{2004A&A.2Haberl}.
Pulse phase resolved spectra for the first three pulsars show that the temperature changes only marginally
with pulse phase. Any model for the pulsed emission from this group of isolated neutron stars
must be able to explain this behaviour.

An important piece of information comes from the detection of broad absorption features in the X-ray spectra 
of several XDINs. At least in the case of \rxb\ the depth of the feature varies strongly with pulse phase.
A likely interpretation of these features is cyclotron resonance absorption which can be expected 
in spectra from magnetized neutron stars with field strength B in the range 
of 10$^{10}$--10$^{11}$ G or 2$\times10^{13}$--2$\times10^{14}$ G if caused by electrons 
or protons, respectively. In the case of \rxb\ the measured $\pdot$, if interpreted as magnetic dipole braking,
rules out the lower range for B, leaving protons or highly ionized heavy atoms as origin.
The measured line energies, summarized in Table~\ref{tab-sum}, then directly provide a measure for B.
The values, assuming proton cyclotron resonance in the magnetosphere of a neutron star with canonical
mass and radius are listed in the last column of Table~\ref{tab-sum}. 
For \rxf\ an independent B estimate of about 1.1$\times10^{13}$ G
can be inferred from the spin-down luminosity required to power the emission nebula \citep{2001A&A...380..221V} 
and the likely age of the star \citep{truemper2004NL}. This is similar to the values derived for the other
stars but may just be outside the range observable at soft X-rays via cyclotron lines.

\begin{acknowledgements}
I thank my collaborators in this project, V. Burwitz, M. Cropper, V. Hambaryan, 
G. Hasinger, R. Neuh\"auser, C. Motch, A. Schwope, J. Tr\"umper, R. Turolla, S. Zane
and V.E. Zavlin. The XMM-Newton project is supported by the Bundesministerium f\"ur 
Bildung und For\-schung / Deutsches Zentrum f\"ur Luft- und Raumfahrt (BMBF / DLR), 
the Max-Planck-Gesellschaft and the Heidenhain-Stif\-tung.
\end{acknowledgements}

\bibliographystyle{aa}

\end{document}